# SUITABILITY OF QUANTIZED DEVS-LIM METHODS FOR SIMULATION OF POWER SYSTEMS

Navid Gholizadeh  
navidg@email.sc.edu

Joseph M. Hood  
hood@email.sc.edu

Roger A. Dougal  
Dougal@cec.sc.edu

**Abstract**

Suitability of the QDL method for analyzing the performance of ac power systems has been evaluated by application to a microgrid. The QDL method is based on a combination of Quantized State Systems (QSS) methods and the Latency Insertion Method (LIM). The accuracy and computational intensity of QDL simulations were evaluated relative to an industry-standard reference method. The key advantages expected of the QDL approach -- including high computational efficiency when the system is operating in steady-state and, when not in steady-state, the need to update only those states that have been affected by quantum level changes of connected states. The expected advantages were largely realized, but with some complications that remain unresolved and require further research such as limit cycle oscillations that emerge in some states after disturbances that should have returned to stationarity. Also, the method in its current state may not be feasible for fault analysis because of the low computational efficiency that results when large disturbances (e.g. faults) create large excursions simultaneously in many system states. The relative strengths and weaknesses of the method are discussed, and some improvements to the method are proposed to overcome the weaknesses. The revelation of observed problems is intended to inspire additional research to overcome those problems.

**Acronyms**

**QDEVS:** Quantized Discrete Event System Specification

**QSS:** Quantized State System

**LIQSS:** Linear Implicit QSS

**LIM:** Latency Insertion Method

**QDL:** Quantized DEVS-LIM Method

**Definitions**

**Computational Intensity**: the number of updates of a QDL atom per unit of simulated time

**Normalized RMS Deviation:** For the *i*th state variable, the RMS difference between values computed by the reference method and the QDL method, with the QDL output resampled at the time step of the reference simulation, and normalized to the dynamic range of the reference simulation over the N samples:

$$D_{rms,\,normalized} = \frac{\sqrt{\sum_{j=1}^{N} \frac{(y_{ij} - q_{ij})^2}{N}}}{\max(y_i) - \min(y_i)} \qquad Eq.\ 1$$

Where *j* is the index of time series.

**Maximum Absolute Percentage Deviation:** For the *i*th state variable, the largest absolute percentage deviation between the value computed by the reference method and the QDL method over the time series of N data points:

$$D_{max} = \frac{\max|y_{ij} - q_{ij}|}{y_{ij}} \times 100\% \qquad Eq.\ 2$$

**Reference solution**: the time-domain solution of a system model (described by a set of DAEs) obtained by use of a 5$^{th}$ order implicit Runge-Kutta method of the Radau IIA family as described in [10] and commonly used for power system solutions.

**Quantization step size**: A constant that defines the set of discrete values the output of a state variable may assume. The output value of a state variable is $\boldsymbol{\Delta Q} \cdot \boldsymbol{k}$, where $\boldsymbol{\Delta Q}$ is the quantization step size, and $\boldsymbol{k}$ is an integer ($\boldsymbol{k} \in \mathbb{Z}$).

**QDEVS**: Quantized Discrete Event System Specification

**LIQSS**: Linear Implicit QSS. Uses a linear approximation of the future state derivatives to predict time to reach next quantized state.

**QDL Atom**: A computational unit (or programming object) that stores and updates the continuous value of a single state variable, and its quantized output value.

**QDL solution**: The time-domain solution of a system model obtained by applying the QDL method

# 1. Introduction

For decades, power system simulation has been accomplished using well-established modeling and simulation methods such as State Space, Modified Nodal Analysis (MNA), and time-slicing numerical integration algorithms. Recent developments in Quantized Discrete Event System (QDEVS) methods, however, appear to present opportunities to innovate power system simulation. In [1] the authors reported how Quantized State System (QSS) methods [2] could be combined with the Latency Insertion Method [3] to solve otherwise intractable problems in electric circuits. The purpose of the research reported here is to evaluate the suitability of those combined methods, abbreviated as QDL (Quantized-DEVS with Latency Insertion), for analyzing the performance of electric power systems, with emphasis on accuracy and efficiency when simulating both slow and fast transient phenomena following common system perturbations such as changes in load level or control setpoints.

The use of the quantized state methods for simulating power systems is motivated by these facts:

1) Some types of power systems often run for long times in near steady state conditions. In these steady states, quantized states should require few or zero updates, so a QDL simulation of the system should be able to advance rapidly through time in between disturbances.

2) If some parts of a system do not remain steady, only the states that move enough to change their quantized outputs will cause changes to propagate to other parts of the system. Stationary states whose inputs do not change should remain relatively stationary.

3) Choice of quantizer size may permit trade-offs between simulation speed and accuracy.

This study itself was further motivated by a desire to discover the types of problems that might be encountered when applying the method so that the community of interested researchers can offer solutions to the problems and ideas for further improvement.

The reference system for this study was a small three-phase ac electric power network having the following components:

1) A generator set, comprising a turbine engine with a speed governor driving a synchronous machine having a controlled exciter,

2) A set of three loads, including an induction motor, a constant impedance ac load, and an ac/dc converter supplying a resistive load, and

3) Several power cables that connect the three-buses of the electrical network.

The reference power system exhibits a range of slow and fast dynamics that effectively exercise the QDL method. The slowest dynamics are associated with the mechanical states, the fast dynamics are associated with electrical states, and speeds of electro-magnetic and magneto-mechanical dynamics fall in between. In order to realize the benefits of the QDL method, the sinusoidally-varying physical system variables were transformed into a rotating reference frame

(e.g. Park transformation [21]) so that the system's state derivatives can be zero in steady-state. The set of system equations are stiff, which poses challenges for conventional time-slice simulation methods with respect to efficiency and numerical stability but presents an opportunity for the strengths of the QDL method to be demonstrated.

## 2. Background

Key concepts of the QDL method are repeated here for completeness, and because ref 1, a conference publication, may not be widely available. QDL is a novel method for time-domain modeling and simulation that combines the principles of Quantized State System (QSS) family of integration methods, Quantized Discrete Event Systems (QDEVS) specification, and the Latency Insertion Method (LIM) modeling approach. LIM allows one to cast a power system model into a QDEVS form. Once a system is described in terms of the QDEVS specification, it can be solved using various QSS integration methods.

### 2.1. Quantized State System Methods

Quantized states are important to achieving our simulation objectives. Any variable whose movement is smaller than its output quantization size will not induce new updates in states connected to it. This contributes to low computing load and fast advancing of simulation time. Discrete event methods are likewise important. When new events are sparse, simulation time can rapidly advance. The well-known Quantized Discrete Event Specification (QDEVS) [4] provides a useful framework for our research. Here we reiterate a few of the principles of QDEVS to put our work into perspective. The Quantized State System (QSS) methods are a series of integration methods based on the QDEVS specification, and described in [4]. These QSS methods provide a QDEVS-compliant way to simulate continuous systems.

The QSS approach assumes that a generic continuous State Equation System

$$\dot{x} = f(x(t), u(t)) \qquad Eq.\ 3$$

can be approximated by a Quantized State System (QSS) in the form

$$\dot{x} = f(q(t), u(t)) \qquad Eq.\ 4$$

where $q$ is a quantized state vector that follows piecewise constant trajectories and is related to the state vector $x$ by the quantum size $\Delta Q$.

References [13][14] define the structure and implementation of atomic DEVS models and general-purpose simulators for QSS systems. The QSS approach guarantees a bounded error[5], so analytically stable systems cannot become numerically unstable when being simulating by a fully-coupled QSS algorithm[14].

The family of QSS integration methods [4] is extensive. The simplest formulation, QSS1, developed in [4], [5], relies on explicit integration and uses first-order estimates of state derivatives to predict the time at which the continuous state $x_i(t)$ will increase or decrease by amount $\Delta Q$ (quantization step size) from the current quantized value $q_i(t)$ to the next higher or lower quantized value. Although QSS1 has some advantages, such as being easy to implement, its disadvantage is that it uses only a first order approximation of the state trajectory to calculate the time to the next event; to get accurate results, $\Delta Q$ has to be relatively small, which produces a large number of steps. QSS2[6] and QSS3[7] use second and third order approximations to more accurately estimate the next event time, however, the computational cost grows with the square root and cubic root (respectively) of the desired accuracy. Furthermore, with stiff systems, these explicit integration methods create fictitious high frequency oscillations[15] that generate large numbers of steps that are costly in computational time and memory size, even when a system is nominally in steady state.

Because we are interested in simulating stiff power systems that include both fast electrical dynamics and slow mechanical dynamics, we chose instead to use the Linear Implicit Quantized State System (LIQSS) methods which were specifically developed to address the concurrent existence of slow and fast dynamics that are inherent to stiff systems [15]. LIQSS methods implement classic implicit integration techniques into the QSS methods. Similar to the way that several variations of QSS methods were developed, so also were variations of LIQSS such as LIQSS1, LIQSS2 and LIQSS3 which perform first, second and third order approximations respectively[15]. The LIQSS2[15], [16], mLIQSS2[15] methods all offer improvements and performance and stability over the original LIQSS1.

Despite the benefits of LIQSS2 or 3 or mLIQSS, two reasons compelled us to use the simpler LIQSS1[15], [16] in this study. First, it was easier to implement the necessary models using LIQSS1, and second, we anticipated that using the first-order method would make it easier to distinguish latency effects from integration effects. If latency methods usefully improve simulator performance for first-order methods, then extensions can later be made to higher-order variations of LIQSS, perhaps with additional gains in performance and stability.

### 2.2 Latency Insertion Method

QSS methods do not intrinsically handle the algebraic constraints that are required to enforce energy conservation in electric circuits. For that purpose, the originators of the QDL method [1] relied on the Latency Insertion Method (LIM) [3]2. The Latency Insertion Method replaces algebraic coupling in the system equations with time latency, while still enabling the benefits of modularization and automatic enforcement of energy conservation constraints that are essential for electric circuits. Traditional power system modeling and simulation tools like general state space models [12] or nodal system model representations, such as the popular Modified Nodal Analysis (MNA) method are well-suited for power system modeling and simulation because they allow to modularize the software and they intrinsically enforce the conservation of energy as expressed by Kirchhoff's voltage and current circuit equations. However, these cannot be directly used with QSS integration methods because they require linear algebra solutions.

LIM is especially valuable for systems containing nonlinear elements which are ubiquitous in power systems. When the number of nonlinear elements becomes large, traditional matrix-based solution methods become inefficient. The latencies required by LIM can be realized in either of two ways: by inserting small, fictitious latency at nodes and branches where negligible physical latency exists, or by exploiting latency (like capacitance or inductance) that naturally exists in system components, but which might otherwise have been eliminated in efforts to simplify a model. LIM permits full node-level system partitioning. To achieve generality, the LIM assumes that Thevenin or Norton transformations can convert any branch or node of a system to this topology. Figures 1 and 2 respectively show how a generic node and a generic branch are represented in the LIM approach.

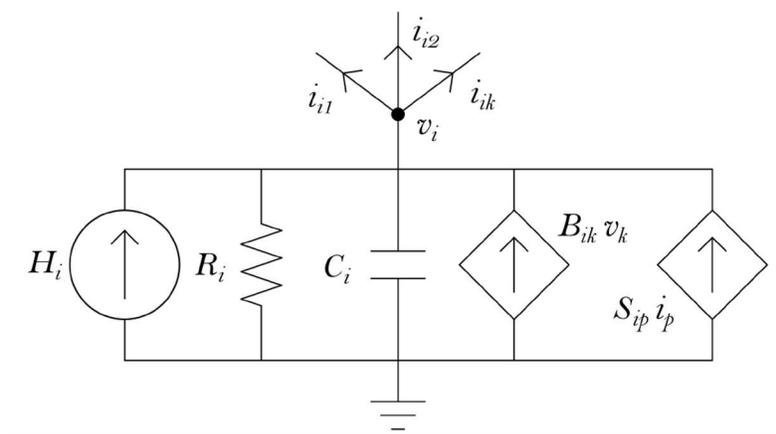

Figure 1: Generic LIM node with dependent sources.

The generic LIM node model shown in Figure 1 includes a voltage-controlled current source (VCCS) and a current-controlled current source (CCCS). These dependent sources provide straight-forward means for modeling energy-conversion coupling circuits, which are common in power systems. $C_i$ is the capacitance at node $i$ and is the element that provides the node voltage latency, $v_i$ is the node voltage, $H_i$ is the current source at node $i$, $R_i$ is the parallel resistance at node I, $B_{ik}$ is the coefficient of the VCCS at node $i$ controlled by the voltage at node k and $S_{ip}$ is the coefficient of the CCCS feeding node $i$ and controlled by current in branch $P$.

The KCL equation for the $i^{th}$ node is:

$$C_i \frac{d}{dt} v_i(t) + G_{ij} v(t) - H_{ij}(t) - B_{ijk} v_k(t) - S_{ijpq} i_{pq}(t) = \sum_{M_i}^{k=1} i_{ik}(t) \qquad Eq.\ 4$$

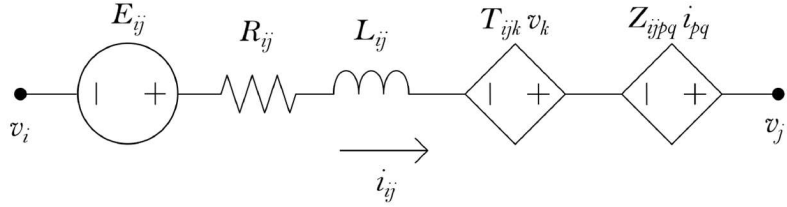

*Figure 2: Generic LIM branch with dependent sources*

The generic LIM branch model with dependent sources is shown in Figure 2 with physical interpretations analogous to those of the LIM node model. The KVL equation from the $i^{th}$ node to the $j^{th}$ node is:

$$v_i(t) - v_j(t) = L_{ij}\frac{d}{dt}i_{ij}(t) + R_{ij}i_{ij}(t) - e_{ij}(t) - T_{ijk}v_k(t) - Z_{ijpq}i_{pq}(t) \qquad Eq.\ 5$$

Where $v_i$ and $v_j$ are the voltages of the connected nodes, $L_{ij}$, $R_{ij}$ and $e_{ij}$ are the series inductance series resistance and series voltage source respectively, and $T_{ijk}$ and $Z_{ijpq}$ are the coefficients of the voltage-controlled voltage source and current controlled voltage source connected to the $k^{th}$ node voltage and the branch current from node $p$ to node $q$ respectively. The LIM method presented here in short is according to the LIM method described in [22]. Note that these parameters, injections and coefficients can, in general, be time-varying and non-linear. According to ref [22] the LIM formulation also allows for externally-controlled, ideal current source branches and voltage source nodes. Although these components are somewhat trivial, they are useful in the derivation of many device models and are worth noting. Because the voltages at the branch ports are effectively dc quantities (as far as the branch model is concerned) during the span of a time step, an ideal source does not require any special consideration by the branch model. Note that ideal voltage source nodes may not be connected in parallel, and ideal current source branches may not be connected in series.

## 3 Reference Power System

The effectiveness of the QDL simulation method was studied using the reference power system shown in Figure 3. The 3-phase, 60 Hz, 4160 V (line-line RMS) power system consists of a generator set source connected to three loads via a network of buses and cables. The generator set is composed of a synchronous machine driven by a gas turbine controlled by an exciter and a speed governor. The loads include an induction motor, a constant impedance (RL) load, and a resistive load behind an ac-to-dc converter.

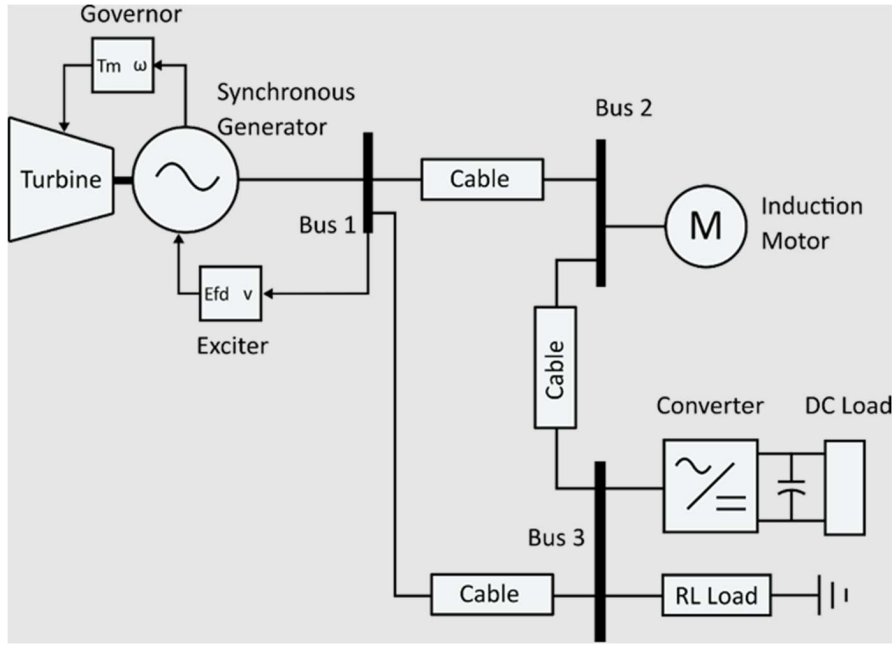

*Figure 3: Reference power system schematic*

The simulation model of the reference system was formulated as follows: The dynamics of the prime mover and the speed governor were combined into a single turbo-governor model. The power converter was described by a non-linear model time-averaged over the switching period. The cables were represented using standard $\pi$ lumped circuit models. All device equations were written in the direct and quadrature coordinates of a rotating reference frame according to the Park transformation which effectively transforms sinusoidal quantities into their phasor equivalents. The system model has 32 state variables, which makes it simple enough to implement using the novel simulation method, yet large and non-linear enough to demonstrate practical utility of the method for analyzing realistic power systems.

The equations of the system model are computed by a set of atoms, where an atom is a computational unit (or programming object) that uses a QDL integration method to update, store, and broadcast a single state variable's continuous internal value and its quantized output value. A model of a single power system component will comprise as many atoms as it has state variables. For each component of the reference system, a model was created compliant to the QDL approach. The simple model of a power cable is shown next to illustrate the approach. QDL representations of the other system components are described in the appendix.

**3.1 QDL Formulation of Cable Model**

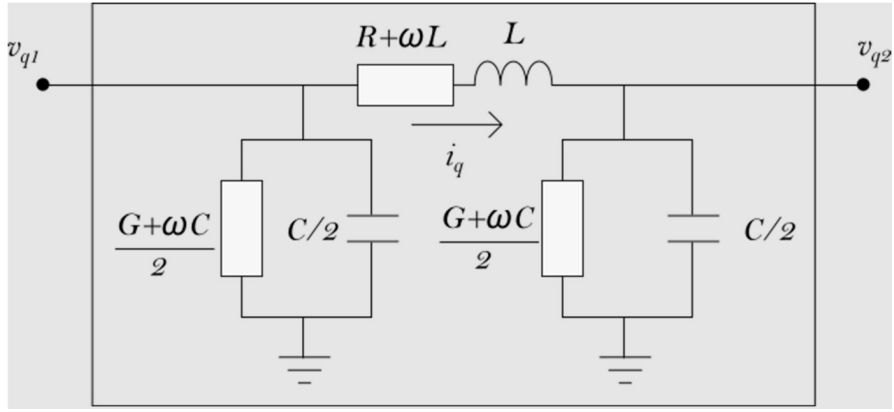

*Figure 4: The LIM branch model for a lumped pi transmission line in the dq frame using a dynamic phasor representation*

The QDL model of a power cable is derived by starting from the traditional lumped π circuit model, which introduces the preferred voltage latency at the end points (the capacitances C/2) and current latency from end to end (the inductance L). Figure 4 shows the circuit equations in terms of the rotating dq coordinates according to the method as described in [18], with separate equations for the direct and quadrature components of the currents and voltages. This yields the following two equations for the q-axis state variables:

$$\frac{d}{dt}i_q(t) = \frac{-(R+\omega L)i_q(t) + (v_{q1}(t) - v_{q2}(t))}{L} \quad \text{Eq. 4}$$

$$\frac{d}{dt}v_{q1}(t) = \frac{-\left(\frac{G+\omega C}{2}\right)v_{q1}(t) + \sum i_{bran}}{C/2} \quad \text{Eq. 5}$$

Where the cable parameters correspond to those shown in Figure 4, and $\sum i_{branch}$ is the sum of currents entering node $q_1$ from all external branches connected to the node. Similar equations describe the other states $v_{q2}, v_{d1}, v_{d2}$ and $i_d$.. Models for all other system components were similarly developed, with some details presented in the appendix, then assembled into the system model to be used in the experiments.

## 4  Simulation Experiments

The electrical connections between the components in the system model were managed programmatically in the simulation framework by mapping QDL node ports to the connected QDL branch ports and summing together the series LIM model contributions connected to a particular system node.

The accuracy and performance of the QDL method were assessed by comparison to a benchmark solution. Conveniently, the QDL formulation of the power system can be used to provide the full state-space description of the system, including the system Jacobian matrix. Therefore, it was possible to run a benchmark simulation by integrating the system equations using a standard time-slicing integration method, specifically a $5^{th}$ order implicit Runge-Kutta method of the Radau IIA family described in [10], a method well-suited for very stiff, non-linear systems. A fixed time step was chosen, with a size appropriate to easily accommodate the fastest eigenvalues of the system.

Accuracy and performance of the QDL method was assessed by exercising the reference power system model through scenarios involving two different types of disturbances. The first disturbance was a load increase, implemented as a 20% step increase of the active power consumed by the RL load on bus 3. The second disturbance is a change of control input, implemented as a 2% increase of the voltage control setpoint of the exciter, from 1.00 per unit to 1.02 per unit. Simulation accuracy was assessed by comparing the deviation of the QDL solution from the reference solution and simulation performance was assessed by comparing the number of state updates required for the QDL solution to the number of timesteps in the reference solution. The performance assessment was necessarily a coarse measure; it compared the performance of new, far-from-optimized code to mature solver code, and there was not a one-to-one alignment of the metrics, but at least the results shed some light on the computational performance that is likely to be achievable with the QDL method.

In general, the state trajectories computed by the QDL method were found to agree quite well with those computed by the reference simulation. The QDL method was expected to generate more frequent computing events immediately following a perturbation, and then less frequent events as the system approached a new steady state. This was sometimes found to be true. The QDL method was expected to update some variables at rates much different than for other variables, and this was also found to be true. But troublingly, the values and update rates of some variables were sometimes inconsistent with the reference simulation and inconsistent with our performance expectations. The expected and the unexpected behaviors will be explained in detail next.

## 5  Results of Simulation Experiments

In the first scenario, at $t = 1$ second, the active power of the RL load was increased by 20 percent. Figure 5 shows the QDL solution of the speed of the synchronous machine ($\omega_r$), and the direct axis components of the machine's current and voltage ($i_{ds}$, $V_{ds}$) in comparison to the quantities computed by the benchmark method prior to and just after the perturbation. For all three of these variables, the quantities computed by the two methods are nearly indistinguishable. In each graph there also appears a dotted line that shows the cumulative number of updates of the corresponding quantized state variable in the QDL solution. The scale of cumulative updates is the same in every graph so that one can immediately see that each variable has a unique update rate.

Several observations can be drawn from the QDL results corresponding to Figure 5:

1) Immediately after the perturbation, the machine speed exhibited a prominent damped sinusoidal oscillation at a frequency of about 16.5 Hz (period about 0.06 s) having initial amplitude about 0.3 rad/s. The QDL solution very accurately tracked the reference solution in respect to the amplitude and phase of this speed oscillation.

2) For all three state variables, the largest absolute percentage deviation between the QDL result and the reference result was very small -- less than 0.0125%. This was the hoped-for result. This indicates that the QDL method is viable and lends confidence that the method was correctly implemented.

3) In the interval before the perturbation, zero state updates were generated for each of the three system variables. The system started in a perfect ac steady state, and it remained there, as expected, until the disturbance occurred. The computational cost of evaluating this time segment was essentially zero. This is exactly the benefit we hoped to achieve.

4) Beginning immediately after the perturbation at 1 second, the slope of the cumulative update line for each variable increased as each state variable moved through many quantization levels. Different states moved through different numbers of quantization levels. The electrical quantities $i_{ds}$ and $v_{ds}$ experienced approximately ten times more updates than the mechanical quantity $\omega_r$. These were all as expected, considering the quantization sizes and the relative motions of these variables.

5) The number of updates of a state variable depended on the quantization size of that variable. For example, $v_{ds}$ experienced about 32,000 updates between 1 and 2 seconds, which is consistent with a 30 V motion and a quantization size of 0.001 V, while $\omega_r$ experienced ~2000 updates with a quantization size of 0.0001 rad/s, which is also consistent with a total motion of approx. 0.16 rad/s.

6) Long after the perturbation, as $\omega_r$ reached a new steady state, its update rate became smaller than immediately after the perturbation, but not zero. The reduction in slope of the update rate was as expected.

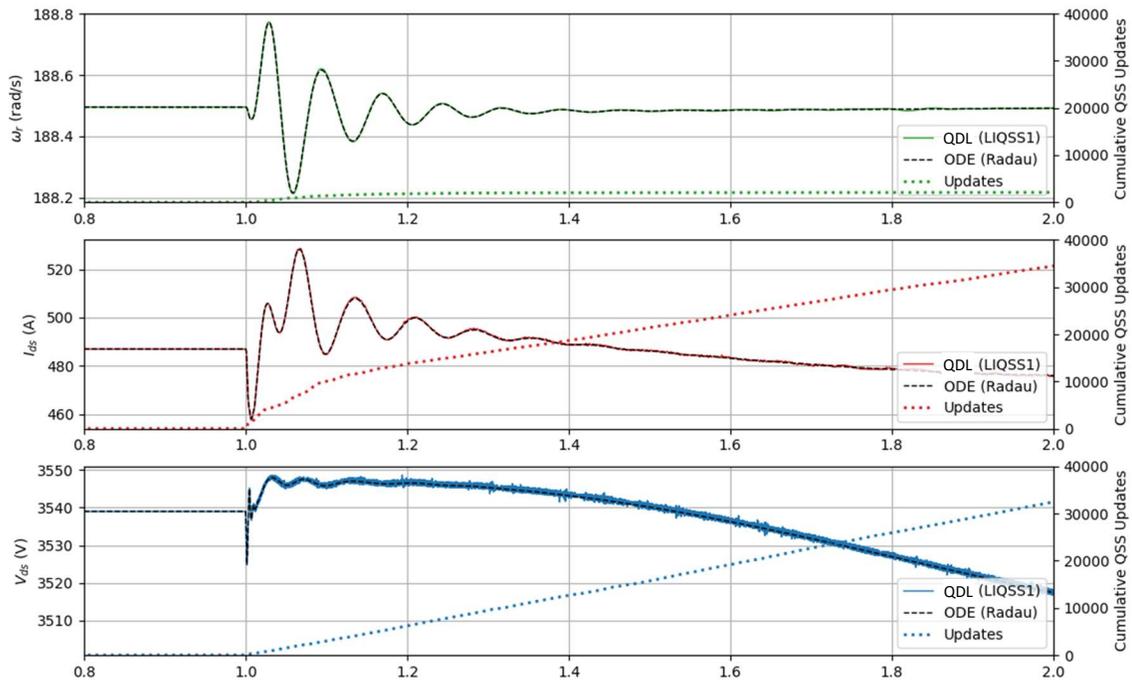

*Figure 5: From top to bottom, rotational speed of the induction machine, direct component of the current of the induction machine, direct component of the voltage of the induction machine respectively, prior to and after the increase of the active power consumed by the RL load.*

Figure 6 shows the cumulative number of atom updates for several of the system states. From this graph, a number of additional conclusions can be drawn.

7) For many of the atoms, the number of update events was significantly less than the number of time steps in the reference solution (labeled as ODE time steps), Notably, for the induction machine speed, the number of updates was smaller by more than two orders of magnitude.

8) For most other atoms, the number of updates was approximately one order of magnitude smaller than the number of ODE time steps.

9) For the fastest-moving variables such as the quadrature axis component of current on cable 2-3 the number of updates was slightly more than the number of ODE time steps but note that the computational cost of updating these individual atoms was very small.

10) If one approximates the computational cost of the QDL simulation by the sum of the update events for all atoms, and if most atoms have update rates one order of magnitude smaller than the ODE solution, and with the total number of system states being of order 10, then the computational costs of the two methods are similar. This is noteworthy given that the QDL algorithms are far from optimized while the reference algorithm is extremely mature.

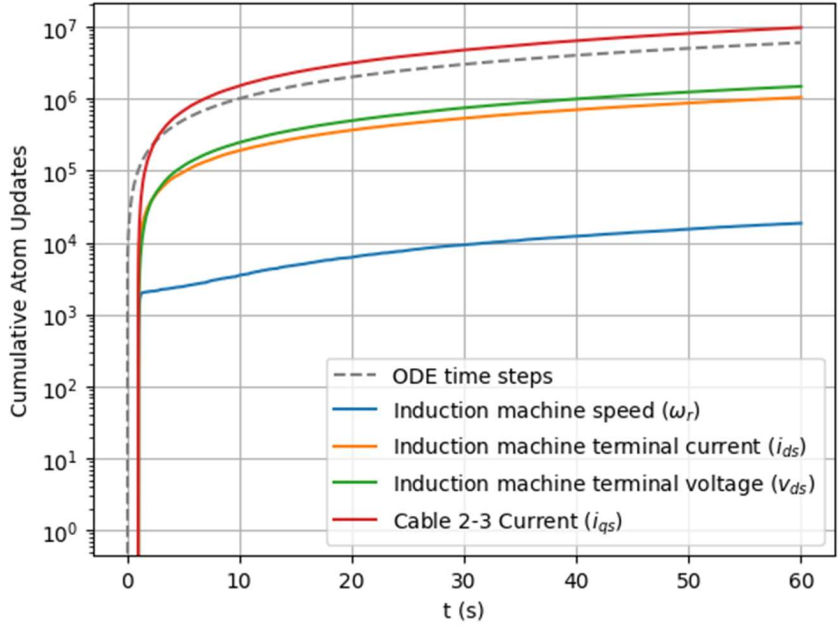

Figure 6: Cumulative QDL atom updates vs. simulation time, induction machine states

In the second scenario the system was perturbed by increasing the voltage reference of the voltage regulator on the generator ($V_{ref}$) by 2% from 1 per unit to 1.02 per unit at t = 1 second. This perturbation of the setpoint created a smaller and less-abrupt dynamic event in the power system compared to the step increase of load that was invoked in the first scenario. Figure 7 shows the same system variables as were plotted for the first experiment. The following additional observations can be drawn:

1) The machine speed exhibited a damped oscillation at the same 16.5 Hz system frequency as was seen in Figure 5, but with much lower amplitude – less than 0.01 rad/s (note the finer resolution on the vertical axis of Figure 7 compared to Figure 5). The amplitude computed by the QDL method was slightly larger than that computed by the reference method. A lower frequency mechanical mode at 0.4 Hz is also evident, about half-a-period of which can be seen in Figure 7, accurately tracked the reference solution with an amplitude of about 0.01 rad/sec. (This 0.4 Hz mode was not evident in Figure 5 due to the coarser scale of the speed axis. More details of this mode will be evident in subsequent figures.) These features were generally as expected, but the reader should pay attention to the slightly higher amplitude of the 16.5 Hz mode that was computed by the QDL method, as this mode will later in the paper be associated with unexpected behavior.

2) The time evolutions of the three plotted system states were consistent with those of the reference solutions. The maximum deviation over the simulation duration was again less than 0.001%. This is as expected.

3) The cumulative update counts of the current and voltage variables at the end of two seconds were roughly the same as in the first experiment, which is consistent with the movements of those state variables through roughly the same range. The cumulative update of synchronous machine rotor speed $\omega_r$ was only 1/10 of the count compared to the first scenario, which is also consistent with the smaller range of the speed response in the second scenario.

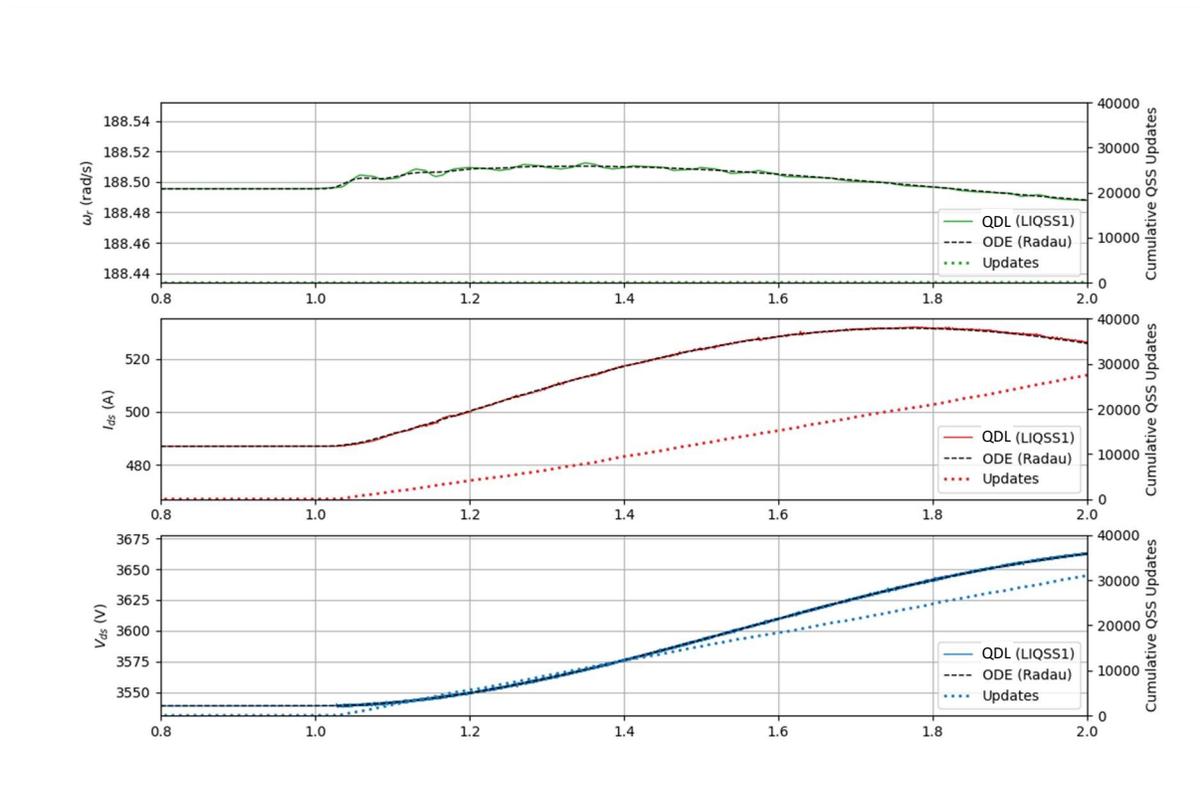

Figure 7: Induction machine speed, direct component of the current and direct component of the voltage respectively, after the AVR voltage setpoint was increased by 2% at 1.0 second (zoom to 0.8-2.0 seconds).

Despite the excellent performance demonstrated thus far, the experiments also revealed at least three interesting problems. A first problem was that some system variables never returned to a quiescent condition when the system should have reached a new steady state. Instead, persistent low-amplitude narrow-band oscillations developed, which caused persistent state update events that should not have been necessary. This diminished the expected high computational efficiency in what should have become a new steady state. A second problem was that seemingly random noise (narrowband but random amplitude) occurred in some system variables after the initial perturbation, even though the underlying (longer term) behavior was correctly computed. A third problem was that the noisy behavior of some state variables obscured the underlying behavior, even when the underlying behavior was correct and could be recovered by low-pass filtering. Each of these three unexpected problems will be explained in more detail next.

## 6 Interesting Problems

Figure 8 shows the speed of the induction machine ($\omega_r$) after the step increase of the RL load, but at three different levels of time and amplitude resolution in order to make clear some of the coarse and fine details that were not apparent in *Figure* 5Figure 5. In Figure 8, one can see that the initial speed oscillation at 16.5 Hz was followed by a lower amplitude damped speed oscillation at 0.4 Hz. The 0.4 Hz component of the speed oscillation computed by the QDL method followed the reference solution rather well in amplitude and phase, but it seemed to be contaminated with a noise signal that had a center frequency curiously close to 16.5 Hz, which, recalling Figure 5, was the natural frequency of the (correctly computed) prompt speed oscillation. .

In the upper graph of Figure 8, near time t =1 second, it is clear that the 16.5 Hz oscillation is rapidly damped, and that the general character of the speed response computed by QDL after damping agrees well with that computed by the reference simulation. This confirms that the correct prediction of machine speed shown in figure 5 continues not just for two seconds, but at least out to 10 seconds. The middle graph of Figure 8 expands the speed scale to show that the correct long-term behavior of the 0.4 Hz mode oscillation is somewhat contaminated by noise having a frequency around 16 Hz. Comparing the noise amplitude between 2 and 4 seconds to that between 14 and 20 seconds (the bottom graph of Figure 8) the noise amplitude appears to increase with time.

As an aside, the upper graph of Figure 8 also shows with the higher resolution scale of the cumulative updates that the update rate of $\omega_r$ was high (about 2000 updates during a fraction of a second immediately after the disturbance) during the transient response, and much lower after the prompt transient died away but it never returned to zero.

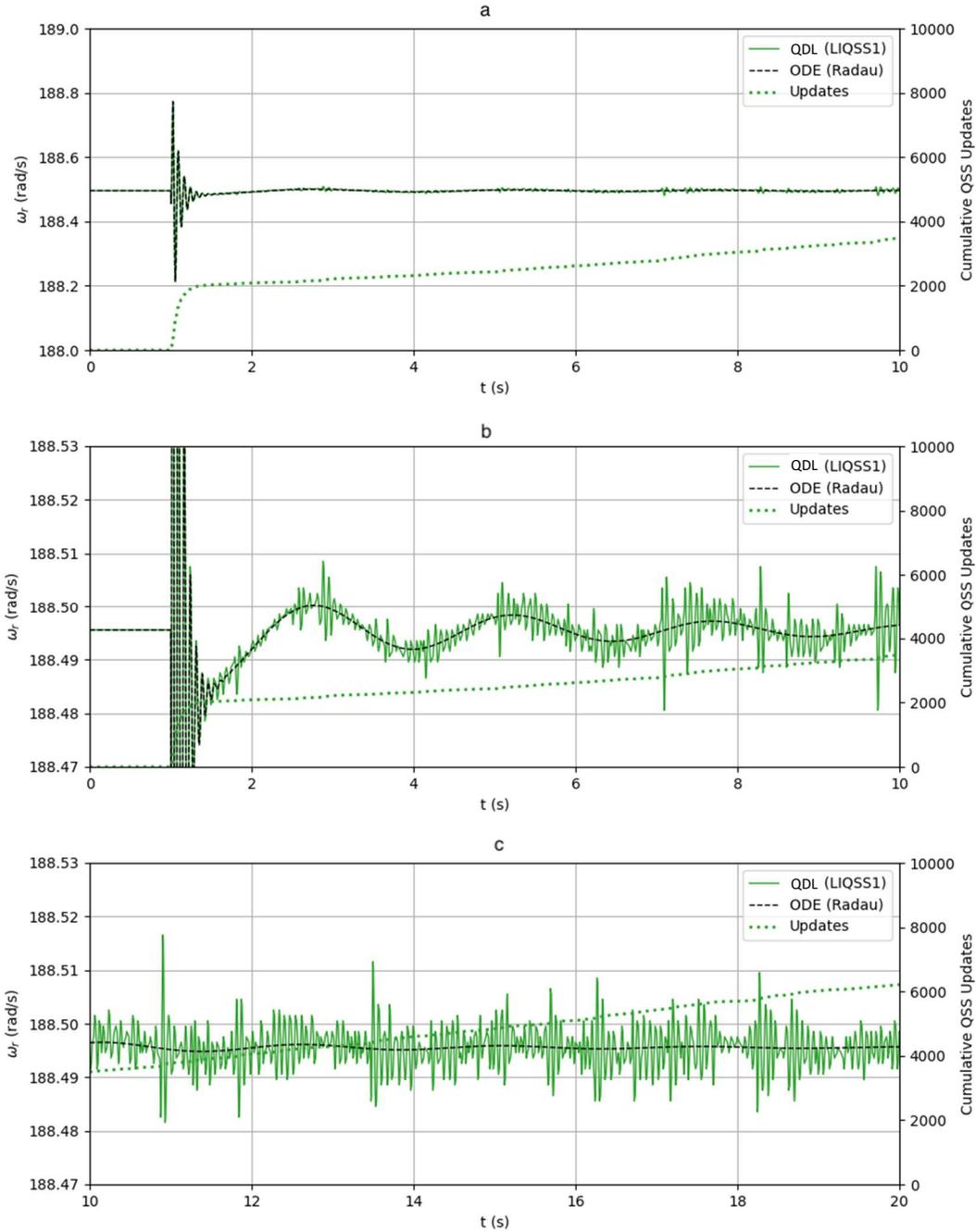

*Figure 8: QDL solution of the speed of the induction machine after load step at t=1s. In upper graph, the first 10 seconds, in middle graph, an expanded vertical scale to show details of the high frequency noise during the first 10 seconds, in lower graph the same vertical scale as middle graph, but showing the QDL solution at later time from 10 to 20 seconds.*

Similar characteristics are evident in Figure 9, which shows the responses after increasing the AVR setpoint. The average of the speed trajectory computed by the QDL method accurately shows the expected amplitude, phase, and damping rate of the 0.4 Hz response, but it is contaminated by the same type of higher frequency noise that was seen in Figure 8, and the amplitude of this noise element seems to grow slowly with time as the system approaches the new steady state. Between t=18 and 20 seconds, the amplitude is the largest. Separate tests (not shown here) revealed that the noise did not grow further at later times but instead reached a limiting amplitude comparable to the amplitude that was seen around t=20s.

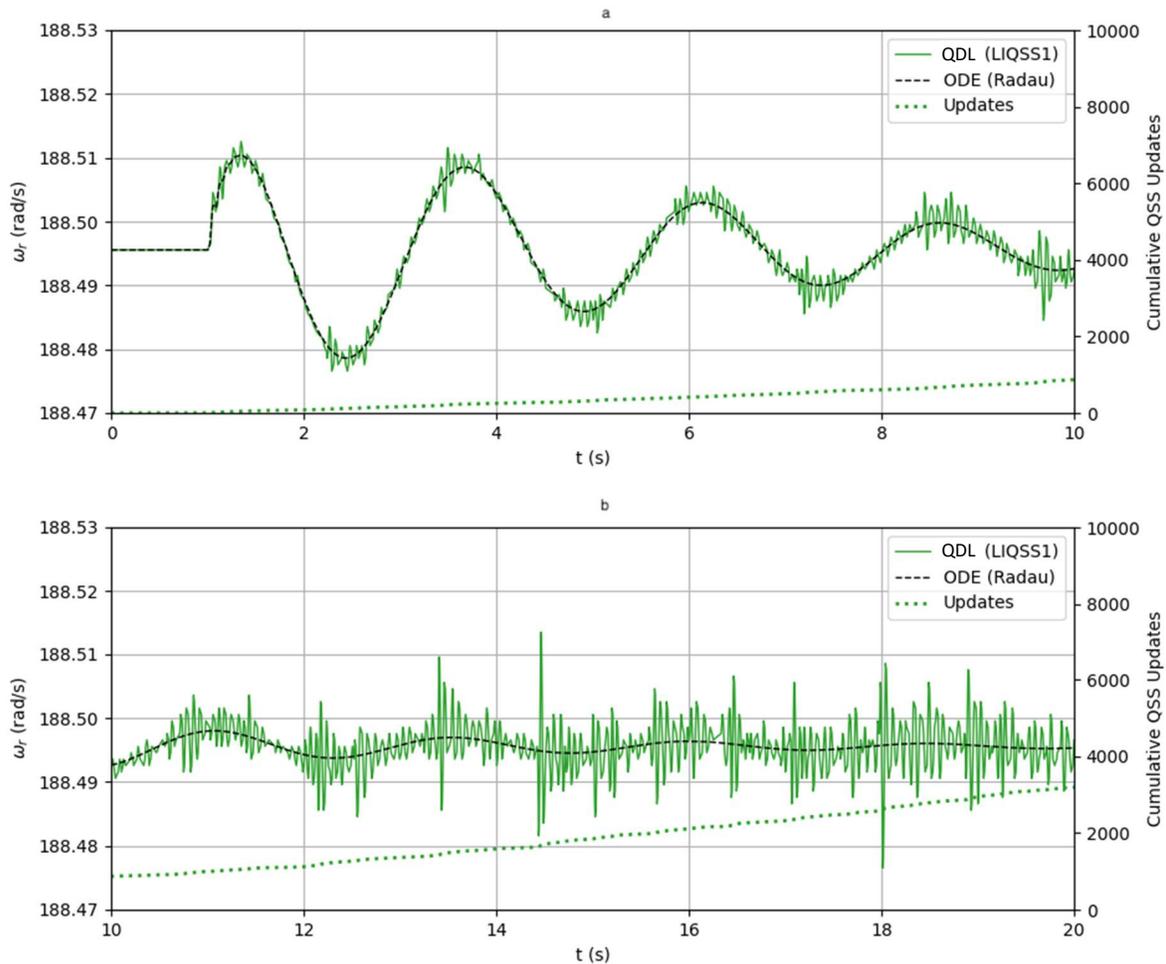

*Figure 9: QDL solution of the induction machine speed. Increasing the AVR voltage setpoint after 1-second by 2 percent. The update count rises steadily after the perturbation. b): QDL solution of the induction machine speed plotted from 10 to 20 seconds. QDL low amplitude oscillations causes the cumulative update line slope to increase steadily.*

Looking at the data presented in **Error! Reference source not found.** and observing that the average of the QDL simulation result appears to coincide with the reference simulation result, one might infer that post-processing of the QDL data with a low-pass filter would yield a result that better tracks the reference simulation. To test this, the QDL data was filtered with a $6^{th}$ order discrete Butterworth low-pass filter having a cutoff frequency of 100 Hz, applied in both the forward and backward directions to cancel the phase shift in order to preserve the transient response as much as possible.

**Error! Reference source not found.** shows a filtered version of the quadrature component of the synchronous machine current ($i_{qs}$) following the perturbation of the first scenario (step increase of RL load), over a sixty second interval, much longer than that shown in the prior figures. The higher-amplitude oscillation at 16.5 Hz that occurred immediately at the time of the disturbance is not quite discernable on this time scale, but it was present, and it was followed by the expected lower-amplitude damped oscillatory response at 0.4 Hz. The QDL solution accurately followed the reference solution for about the first ten seconds. After that, one can see that the solution computed by the reference method became fully damped to a new steady state, whereas the QDL simulation (low pass filtered) continued to oscillate indefinitely at 0.4 Hz. The QDL simulation method apparently introduced a pumping effect at the 0.4 Hz frequency characteristic of the mechanical mode.

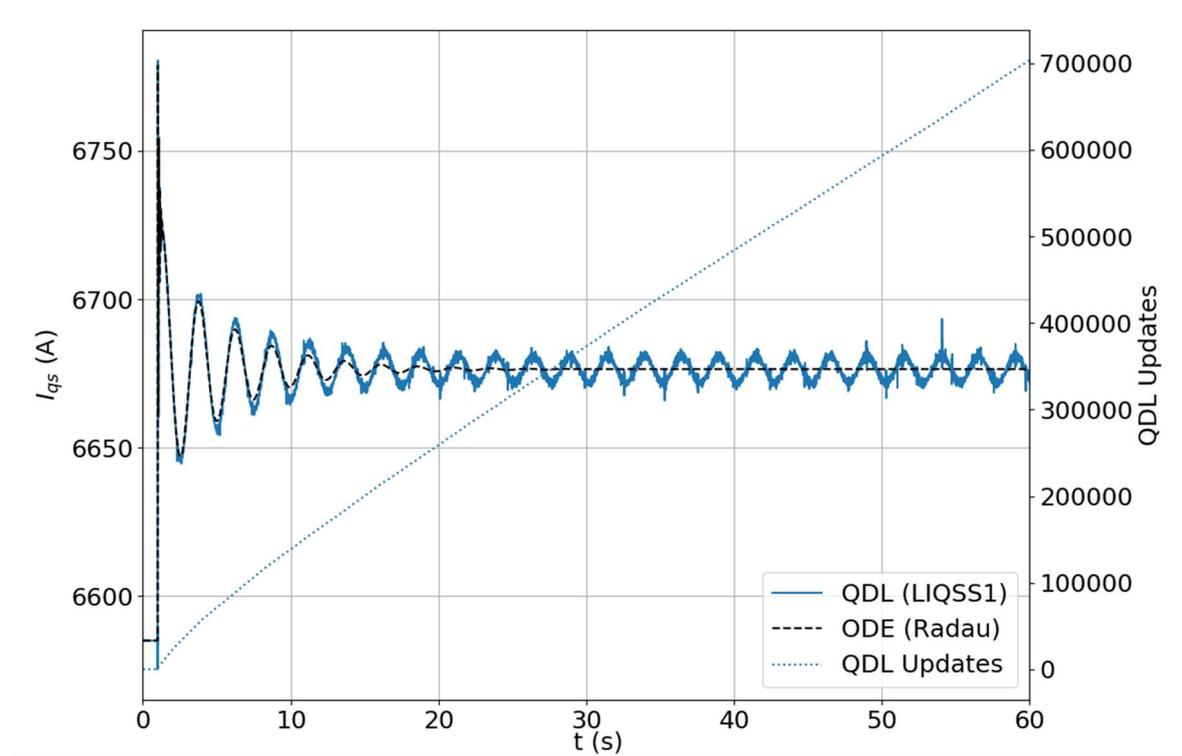

*Figure 10: quadrature component of the synchronous machine stator current (filtered, fc=100 Hz)*

The curve of the filtered QDL solution can be approximated by the analytic function:

$$x(t) \cong A \cdot e^{-\lambda t} \big(\cos(\omega t + \varphi) + \sin(\omega t + \varphi)\big) + x_{dc} + B \cdot \cos(\omega t + \varphi) \qquad \text{Eq. 6}$$

with estimated parameters as $A = 72$, $\lambda = 0.4\ s^{-1}$, $\omega = 2.5$ rad/s, $\varphi = 3.14$, $x_{dc} = 6677$ and $B = 5$. The QDL simulation correctly computed the nominal current of 6677 A and the amplitude and frequency of the decaying oscillations at 30A and 0.4 Hz respectively just after the disturbance, and the correct initial damping rate of $0.16\ s^{-1}$, but the oscillations ultimately failed to die away; $i_{qs}$ exhibited persistent oscillation with amplitude of about 9.96 A at the characteristic frequency of 0.4 Hz. Reference [13] revealed that oscillations of this type seem to have a complex relationship with the quantization step size, the initial conditions, and the nature (amplitude and rate) of the applied perturbation.

The simulation of stiff systems using traditional QSS methods is known to result in high-frequency oscillations of quantized state quantities that should be nominally in a steady-state condition [2][10][11]. Natural frequency is prominent in both the QDL and Reference simulations immediately after the perturbation, but only in the Reference case do the oscillations die out. In the QDL method, steady state oscillations persist indefinitely. The more advanced QSS methods [14][15] should be investigated.

Although this post-processing approach (low pass filtering of the noisy result) might be viable, it is clearly undesirable as it requires the person running the simulation to know which frequencies to keep and which to eliminate, which may not be possible if one has not also run another reference simulation to know which frequencies to eliminate. Obviously, if one already had the reference simulation, it would seem to obviate the need for a model computed by this new simulation method.

While the machine speed computed by the QDL method does correctly track the reference solution in spite of some amount of noise, some other states exhibited such noisy behaviors that the noise obfuscated the underlying average behavior. The worst cases for noisy states were associated with cables, with the biggest deviation being found in the quadrature-axis current of the cable connecting bus 2 to bus 3. We suspect that the cables exhibited the largest deviations because they also had the highest natural frequencies. **Error! Reference source not found.** shows the RMS deviation of system variables, normalized to its own range of motion - a very stringent criteria for accuracy compared to normalizing to the nominal value -- sorted from largest to smallest, averaged over the entire 60 second duration of the simulation. The normalized RMS deviation (defined by eq. 1) was largest in the cables and the AVR system, while the deviations in other state variables were below 4%.

Applying the same low-pass filter, but with a cutoff frequency of 50 Hz instead of 100 Hz, produced the data shown in Figure 12, Figure 13 and Figure 14, each on a different time scale. The whole 60-second behavior of filtering is shown in Figure 12. The filtered version of the QDL solution (Figure 13) tracks the benchmark simulation very well through the initial transient response (although a small filtering artifact (not resolvable here) does appear at the instant of the perturbation). Figure 17 expands the last 5 seconds of the simulation showing the oscillations in the filtered signal. The filtering process reduced the normalized RMS deviation of the QDL solution from 9.30% to 1.02%. It is expected that filtering can be applied to the other states with similar results. It is recognized that this post-filtering approach is not desirable -- it would be better to discover and resolve the underlying cause of the noisy behavior – but it appears to be effective.

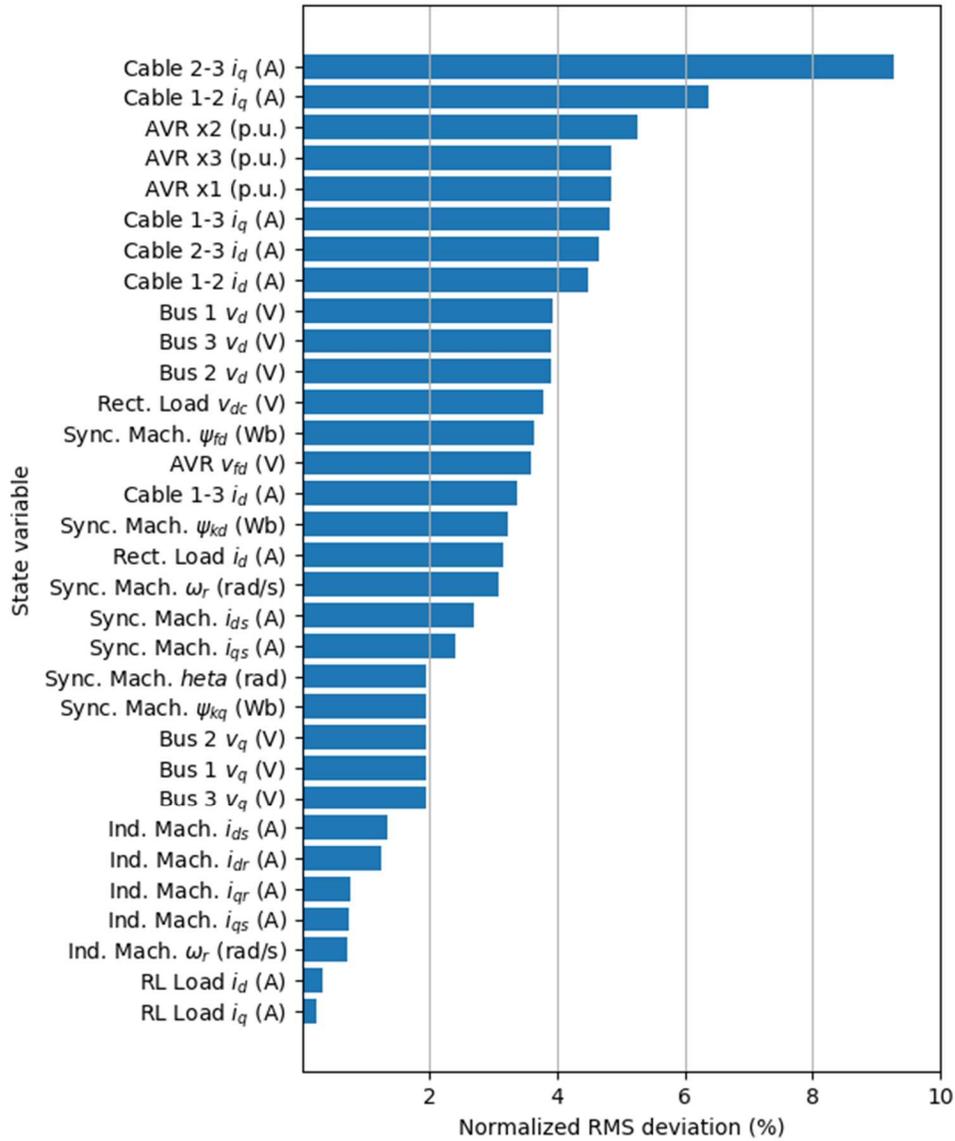

*Figure 11:Normalized RMS deviation of each state variable from the reference solution. (normalized with respect to the dynamic range of motion of that state variable during the simulation*

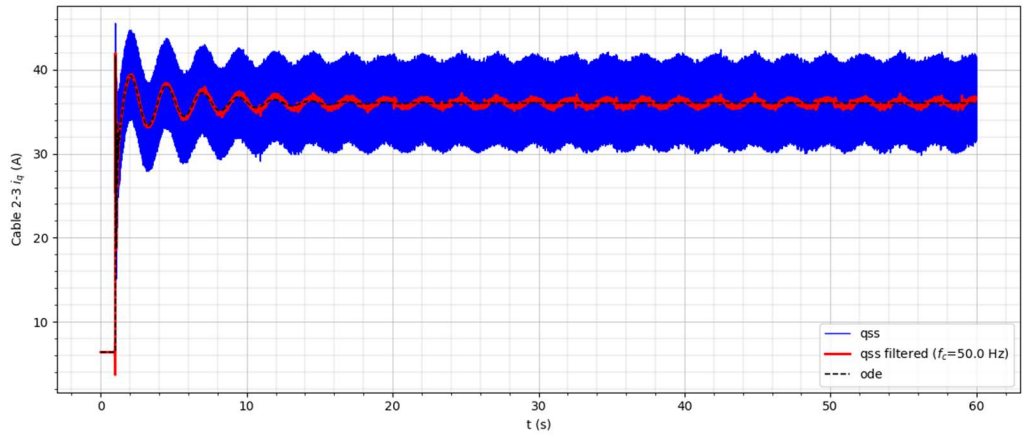

*Figure 12: Q-axis current for cable connecting buses 2 and 3, with QDL solution unfiltered and filtered, and the reference solution, over the entire 60s duration of the experiment.*

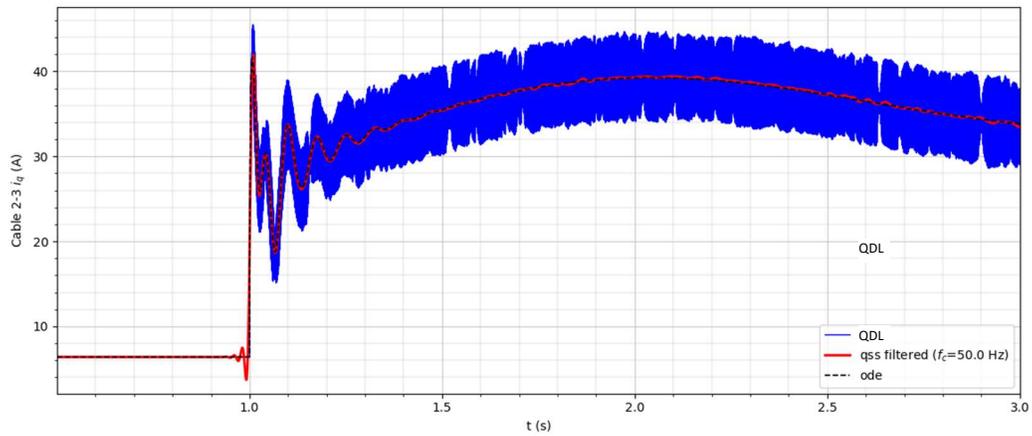

*Figure 13: Q-axis current for cable connecting buses 2 and 3, with QDL solution unfiltered and filtered, and the reference solution, expanding detail in the period from 1 to 3s.*

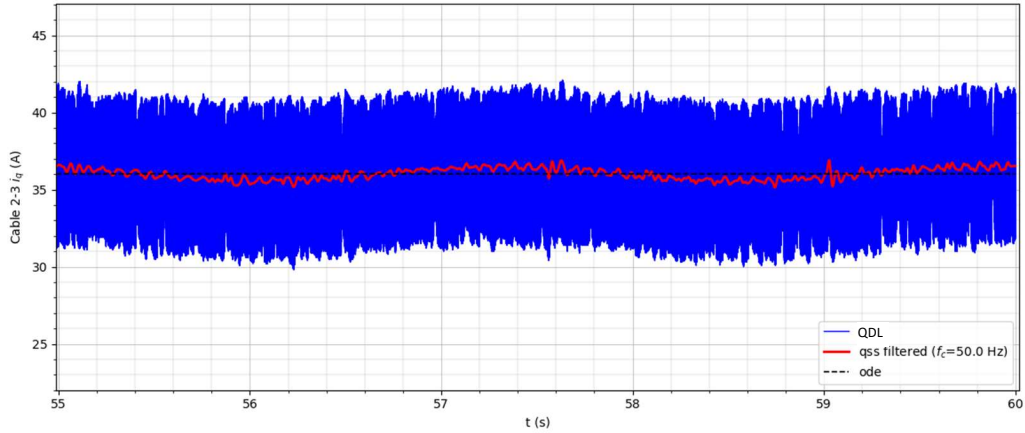

*Figure 14: Q-axis current for cable connecting buses 2 and 3, with QDL solution unfiltered and filtered, and the reference solution, expanding detail in the final 5s of the 60s experiment.*

Finally, we note a general weakness of the QDL method that is not presented here with data; the speed of computing is not good under the condition of large-signal perturbations such as would be induced by a fault analysis. The large excursion of all state variables away from their nominal operating points creates event storms as each variable moves through a very large number of quantized levels. As a result, it was found to be not possible to complete the simulation of a short-circuit fault.

For the QDL method to become fully successful, strategies must be developed to eliminate or compensate the various shortcomings. It was not possible to resolve all of them within the initial scope of this study, but we felt it important to present the problems so that we and others can eventually resolve them. Some problems will require more research than others. For example, it is not expected that even the latest LIQSS methods (such as the modified, $2^{nd}$ order mLIQSS2) will yield better performance when analyzing power system fault scenarios in which event storms are caused by large excursions of many system states. This type of problem will likely require development of new methods such as state jumping. Perhaps the QDL method will eventually prove ineffective for fault analysis simply because its most promising virtue (high efficiency in steady-state) is inherently its biggest liability in the opposite conditions (far from steady state).

## 7 Conclusion

The QDL method using LIQSS1 integrator and quantizer was able to track the moving equilibrium of a relatively complex (32 state) non-linear power system. With the quantization step sizes used in this experiment, some of the state variables were computed to within 0.01% of the values computed by the reference simulation, and most of the state variables were computed to within a RMS deviation normalized to the range of motion, not to the nominal value (much more stringent), of 1% to 3% compared to the reference simulation. Proxy metrics for computational intensity (the number of QDL atom state updates, and the number of timesteps for the reference solution) imply that the theoretical performance of optimized code should eventually result in much faster performance for the QSS-LIM solution than for the reference simulation (the experimental code for the QDL method was far from optimized).

Experiments so far have uncovered three significant issues that must be resolved in order for this simulation method to realize the advantages expected of it; the persistent oscillations that inhibit achievement of a steady state must be eliminated, the high frequency and large amplitude noise on some states must be eliminated, and a way must be found to deal with the discrete event storms that occur during the large state excursions such as those associated with power system fault analysis.

## 8   Acknowledgements

Two of the authors received partial support for this work from the US Office of Naval Research under grants N00014-16-1-2956 and N00014-21-1-2124.

## Appendix

Some of the immediate implementation details and models are presented here. More complete models and plots are available in the GitHub repository associated with this paper.

One could choose to apply a single uniform quantization size to all state variables, but in the system associated with this study, where each state variable has a different magnitude and range of interest, we chose different quantization sizes for some variables based on the dynamic ranges of those variables.

System Device QSS-LIM Model Descriptions

RL load Model:

The series RL load is modeled as two LIM branches in the dq domain connected to ground. Because this model is easily described by two LIM atoms, the equations are not given here. (see LIM atom description)

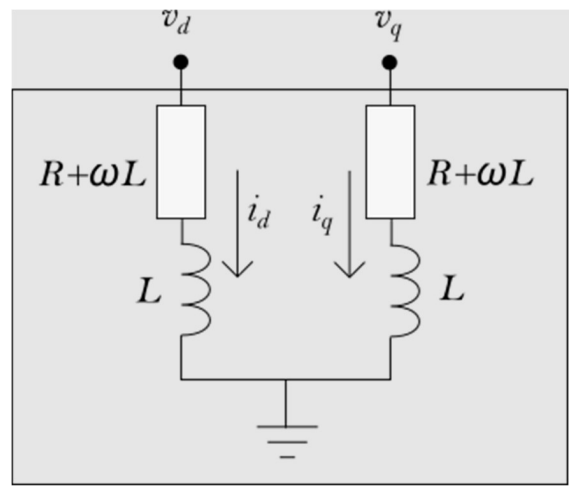

Figure 15: RL load model

Transformer Rectifier Model:

The transformer-rectifier load model is based on the model found in [21], and includes non-linear coupling between the dc side and the ac side d- and q-axis components.

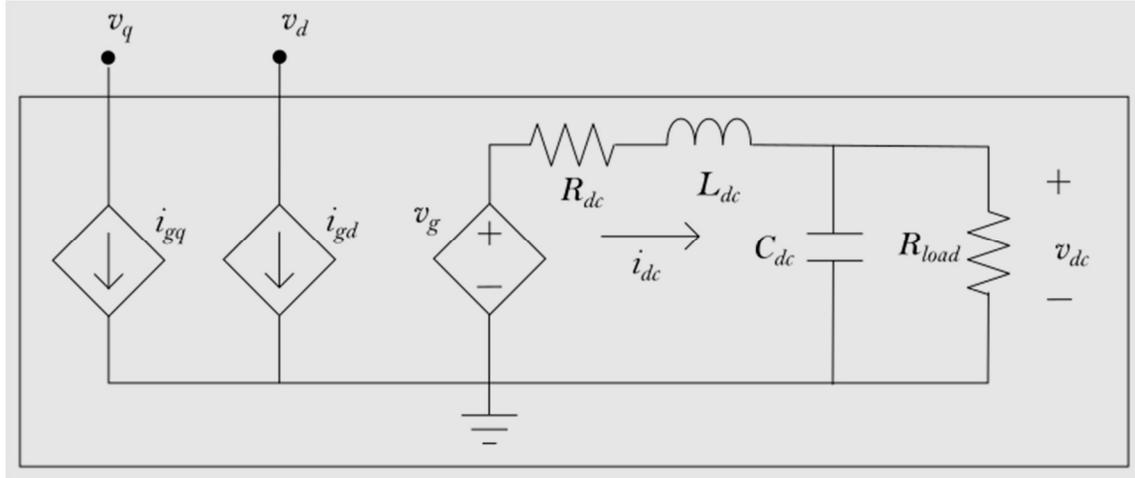

Figure 16:Transformer rectifier model

$$i_{gd} = S_d i_{dc}$$
$$i_{gd} = S_q i_{dc}$$
$$v_g = \frac{v_q}{S_q} + \frac{v_d}{S_d}$$

Where

$$S_d = 2\sqrt{\frac{3\sqrt{3}}{2}\frac{1}{\pi}\cos(\phi)}$$

$$S_q = -2\sqrt{\frac{3\sqrt{3}}{2}\frac{1}{\pi}\sin(\phi)}$$

The dc side of the model is a LIM branch (Figure 2) that is coupled to a LIM node (Figure 1).

Synchronous Machine:

The synchronous machine [21] is modeled in the DQ domain as two LIM branches for the electrical model (one for the q-axis terminal, and one for the d-axis), and a LIM node for the mechanical dynamics.

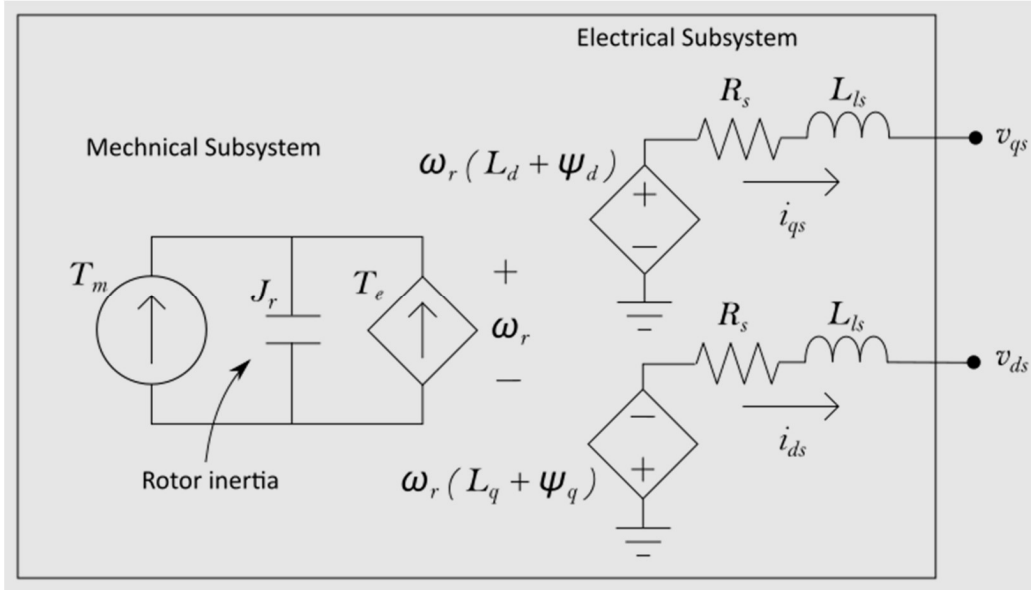

Figure 17:DQ Synchronous machine model (LIM portions only)

The electromechanical dynamics are modeled with the flux equations:

$$\frac{d}{dt}\psi_{kq} = \frac{r_{kq}}{L_{lkq}}(\psi_{kq} - L_q i_{qs} - \psi_q i_{qs} + L_{ls} i_{qs})$$

$$\frac{d}{dt}\psi_{kd} = \frac{r_{kd}}{L_{lkd}}(\psi_{kd} - L_d i_{ds} - \psi_d i_{ds} + L_{ls} i_{ds})$$

$$\frac{d}{dt}\psi_{fd} = \frac{r_{fd}}{L_{lfd}}(\psi_{fd} - L_d i_{ds} - \psi_d i_{ds} + L_{ls} i_{ds} - v_{fd})$$

where the electrical torque $T_e$ is coupled to the terminal currents $i_{qs}, i_{ds}$ as

$$T_e = \frac{3}{2}\frac{P}{2}(\psi_{ds} i_{qs} - \psi_{qs} i_{ds})$$

and the q and d-axis equivalent fluxes and inductances are given by

$$L_q = L_{ls} + \frac{L_{mq} L_{lkq}}{L_{lkq} + L_{mq}}$$

$$L_d = L_{ls} + \frac{(L_{md}L_{l\psi d}L_{lkd})}{L_{md}L_{l\psi d} + L_{md}L_{lkd} + L_{l\psi d}L_{lkd}}$$

$$\psi_q = \frac{L_{mq}}{L_{mq} + L_{lkg}} \psi_{kg}$$

$$\psi_d = \frac{L_{md}\left(L_{md} * \left(\frac{\psi_{kd}}{L_{lkd}} + \frac{\psi_{\psi d}}{L_{l\psi d}}\right)\right)}{1 + \frac{L_{md}}{L_{l\psi d}} + \frac{L_{md}}{L_{lkd}}}$$

and the rotor inertia is defined as:

$$J \frac{d^2 \theta_m}{dt^2} = T_m - T_e$$

Where:

- $\theta_m$ = angle is rad
- $T_m$ = Mechanical toruqe in Nm
- $T_e$ = electrical torque developed in Nm

Turbo-Governor Model

The dynamics of the prime mover and the speed governor are combined into a single model- consisting of electrical LIM nodes and branches and mechanical differential equations which then are translated into QDL atoms- that determines the mechanical torque $T_m$ from the machine speed using the equivalent PI controller constants $K_p$ and $K_i$ as

$$T_m = K_p \delta_r + K_i \theta_r$$

where the speed deviation $\delta_r$ and rotor angle $\theta_r$ (the difference between the synchronous electrical angle and the mechanical angle) are given by

$$\delta_r = \omega_s - \omega_r$$

$$\frac{d}{dt}\theta = \delta_r.$$

**Exciter Model:**

The ac terminal voltage of the synchronous machine is regulated by the exciter. The exciter is described by a simplified IEEE AC8B AVR model, which includes three differential equations for internal signal states, and a fourth differential equation for the output voltage.

The input is the per unit terminal voltage magnitude -$v_{t,pu}$- and the output is the per unit field voltage $v_{fd}$. The differential equations are

$$\frac{d}{dt}x_1 = v_{t,pu} - \frac{1}{T_{dr}}x_1$$

$$\frac{d}{dt}x_2 = x_1$$

$$\frac{d}{dt}x_3 = \left(K_{ir} - \frac{K_{dr}}{T_{dr}^2}\right)x_1 + \frac{K_{ir}}{T_{dr}}x_2 - \frac{1}{T_a}x_3 + \left(\frac{K_{dr}}{T_{dr}} + K_{pr}\right)v_{t,pu}$$

$$\frac{d}{dt}v_{fd,pu} = \frac{K_a}{T_a T_e}x_3 - \frac{K_e}{T_e}v_{fd,pu}$$

where the per unit terminal voltage $v_{t,pu}$ is related to the machine dq quantities and the rated line-line terminal voltage magnitude $V_{base,LL}$ as

$$v_{t,pu} = \frac{\sqrt{v_{qs}^2 + v_{ds}^2}}{V_{base,LL}}$$

and the per unit output of the AVR $v_{fd,pu}$ is then scaled by base voltage $V_{base,LL}$ to produce $v_{fd}$.

$$v_{fd} = v_{fd,pu} \cdot V_{base,LL}$$

Induction Machine model:

The induction machine is modeled in the DQ domain as two LIM branches for the electrical model and a LIM node for the mechanical dynamics. Electrical states are translated into LIM nodes and branches and mechanical states are translated into differential equations and finally all are translated into QDL atoms.

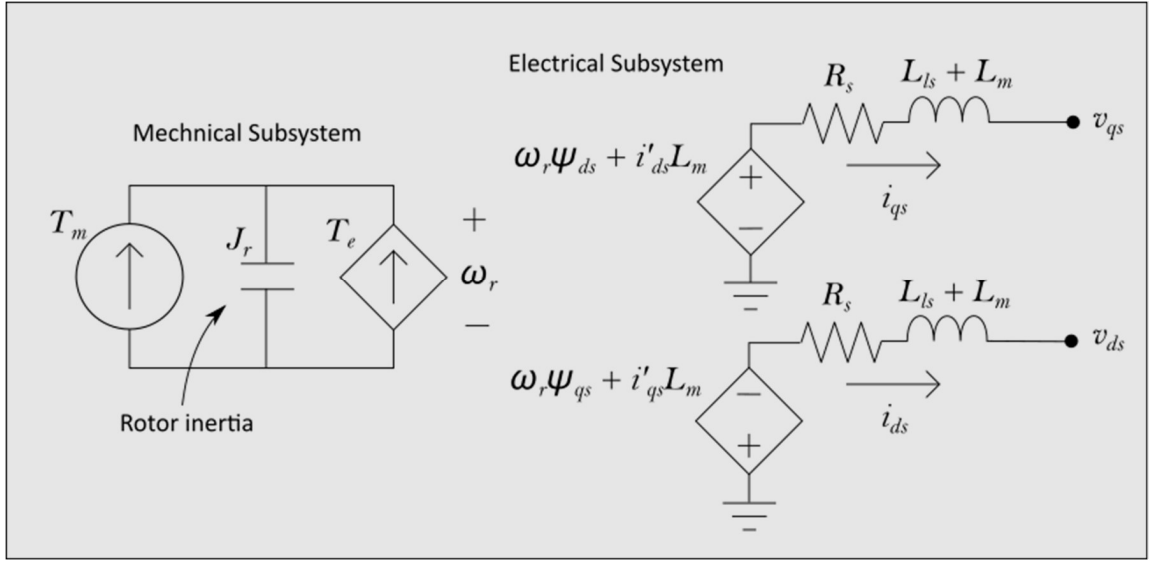

Figure 18: Induction Machine

$$V_{qs} = R_s * i_{qs} + \omega_r * \psi_{ds} + (L_{ls} + L_m) * \frac{d}{dt}i_{qs} + L_m * \frac{d}{dt}i_{qr}$$
$$V_{ds} = R_s * i_{ds} - \omega_r * \psi_{qs} + (L_{ls} + L_m) * \frac{d}{dt}i_{ds} + L_m * \frac{d}{dt}i_{dr}$$
$$V_{qr} = R_r * i_{qr} + (\omega_s - \omega_r) * \psi_{dr} + (L_{lr} + L_m) * \frac{d}{dt}i_{qr} + L_m * \frac{d}{dt}i_{qs}$$
$$V_{dr} = R_r * i_{dr} - (\omega_s - \omega_r) * \psi_{qr} + (L_{lr} + L_m) * \frac{d}{dt}i_{dr} + L_m * \frac{d}{dt}i_{ds}$$
$$\psi_{qs} = L_{ls} * i_{qs} + l_m * (i_{qs} + i_{qr})$$
$$\psi_{ds} = L_{ls} * i_{ds} + l_m * (i_{ds} + i_{dr})$$
$$\psi_{qr} = L_{lr} * i_{qr} + l_m * (i_{qr} + i_{qs})$$
$$\psi_{dr} = L_{lr} * i_{dr} + l_m * (i_{dr} + i_{ds})$$

These equations were solved for the current derivatives using a symbolic processor, then the derivatives were treated as LIM nodes.

For mechanical side:

$$\frac{d}{dt}\omega_r = \frac{P}{2J}\left(\frac{3P}{4}\left(L_{ls}i_{ds} + L_m(i_{ds} + i_{dr})\right)i_{qs} - \left(L_{ls}i_{qs} + L_m(i_{qs} + i_{qr})\right)i_{ds} - T_b\left(\frac{\omega_r}{\omega_s}\right)^3\right)$$